# Cellular Micromasonry: Biofabrication with Single Cell Precision

S. Tori Ellison[1], Senthilkumar Duraivel[1], Vignesh Subramaniam[2], Kjell Fredrik Hugosson[3], Joseph J. Lebowitz[4], Habibeh Khoshbouei[4], Mark Q. Martindale[3], Thomas E. Angelini[2,*]

[1]Univeristy of Florida Department of Materials Science and Engineering, Gainesville, FL, USA.

[2]Univeristy of Florida Department of Mechanical and Aerospace Engineering, Gainesville, FL, USA.

[3]The Whitney Laboratory for Marine Bioscience, St. Augustine, FL, USA

[4]University of Florida Department of Neurosurgery, Gainesville, FL, USA

*Correspondence to: t.e.angelini@ufl.edu.

**In many tissues, cell type varies over single-cell length-scales, creating detailed spatial heterogeneities fundamental to physiological function. To gain understanding of this relationship between tissue function and detailed structure, and to one day engineer structurally and physiologically accurate tissues, we need the ability to assemble 3D cellular structures having the level of detail found in living tissue. Here we introduce a method of 3D cell assembly that has a level of precision finer than the single-cell scale. With this method we create numerous structures having well-defined spatial patterns, demonstrating that cell type can be varied over the scale of individual cells and showing function after their assembly. This technique provides innumerable opportunities to study cellular behavior in defined contexts including complex interactions operating during embryogenesis.**

## Introduction

The different cell types that constitute living tissue are often structured into highly heterogeneous and complex spatial patterns; cell type can differ over length-scales as small as a single cell within a given tissue *(1, 2)*. This spatial heterogeneity is broadly linked to different types of tissue function. For example, to maintain high rates of molecular exchange in the liver, a network of endothelial cells, called the sinusoid, permeates the periportal zone where every hepatocyte can be found within one or two cell diameters of an endothelial capillary *(3, 4)*. Another dramatic example is found in the pancreatic islet, where the five main cell types of the islet are located within a few cell diameters of one another, making frequent contacts with acinar and ductal cells of the exocrine pancreas *(5, 6)*. This highly heterogeneous arrangement of cell types within the islet's relatively small volume allows cells to communicate through secretion and maintain blood glucose homeostasis *(5, 6)*. The connection between small-scale structural heterogeneity and tissue function is also exhibited by glandular acini *in vitro (7-9)*. These hollow spheres are made from single epithelial monolayers surrounded by a basement membrane; the signaling between the basement membrane and cell nuclei is crucial for acini to develop and function *(7, 10-12)*. While glandular acini represent an *in vitro* system in which the link between tissue structure and function can be studied in detail, it remains exceedingly challenging to reproduce the complex cellular patterns found more generally *in vivo*. For example, relying on spontaneous or guided processes of multi-cellular self-assembly within bioengineered tissues is time consuming and does not precisely reproduce the detailed structural and functional heterogeneity at the single-cell scale found within *in vivo* tissues *(13, 14)*. 3D bioprinting provides control and repeatability for structuring *in vitro* tissue models, but current tools are not sufficiently precise to produce spatial variations in cell type over the scale of even a few cells, much less a single cell. To create tissue models that reproduce the spatial heterogeneities found within *in vivo* tissue, new biofabrication tools with single-cell precision are needed. Without such tools, our basic understanding of how tissue function collectively emerges from spatially heterogeneous tissue structure will continue to depend on observations and methods that rely dominantly on cell-directed organization, which have persistently challenged researchers.

Here we introduce a method that enables us to emulate the heterogeneous spatial patterning found *in vivo* with single-cell precision. The method, which we call "cellular micromasonry," combines a soft 3D support medium with micromanipulation and 3D microscopy. The 3D support medium is a phase of soft matter made from jammed granular-scale microgels – hydrogel microparticles packed together that form the microscopic equivalent of the "ball pit" children play in *(15-17)*. Children in ball pits can lay still, supported by the static forces of the packed balls, yet they can also swim through the balls, embedding themselves deep within their surroundings. By analogy, here we use micromanipulators to grasp, translate, and place cells in "ball pit" made from microscopic hydrogel particles swollen in liquid cell growth media (Fig. 1A). This microgel medium is stiff enough to support the cellular structures, but soft enough that a microcapillary holding cells can easily be translated through it; the microgels' low polymer concentration limits the physical stress that cells experience as we build with them *(15-18)*. We use this method to methodically place cells in patterns that comprise heterogeneous populations, controllably alternating between different cells, one-by-one. With this approach, we demonstrate that the highest degree of spatial heterogeneity can be achieved – single cell precision. To test for function, we study molecular transport through gap junctions, observing calcein dye diffusing from cell-to-cell, and we show that glandular acini can develop within this medium. This cellular micromasonry method enables the building of stratified, precise cellular structures for detailed investigations of the relationship between structure and function in models of both developing and mature tissues.

**Results**
To create precise cellular structures, we harvest cells from their traditional culture conditions and manually disperse them with a pipette into the microgel culture medium contained in a glass-bottomed petri dish (see Methods for cell types and culture details). The dish is mounted onto a temperature-controlled stage atop an inverted laser scanning confocal microscope. When using carbonate-based culture buffers, humidified $CO_2$ is gently blown onto the sample surface to maintain neutral pH in the microgel culture medium. Using the microscope, we identify a chosen cell, translate the tip of a microcapillary to its surface using the micromanipulator, and lightly aspirate using a CellTram (Eppendorf), applying suction. Once the cell is captured, it is translated to the desired location and deposited. By repeatedly capturing, translating, and depositing cells, we assemble structures suspended in 3D space without having to build up from a solid support (Fig. 1B-F).

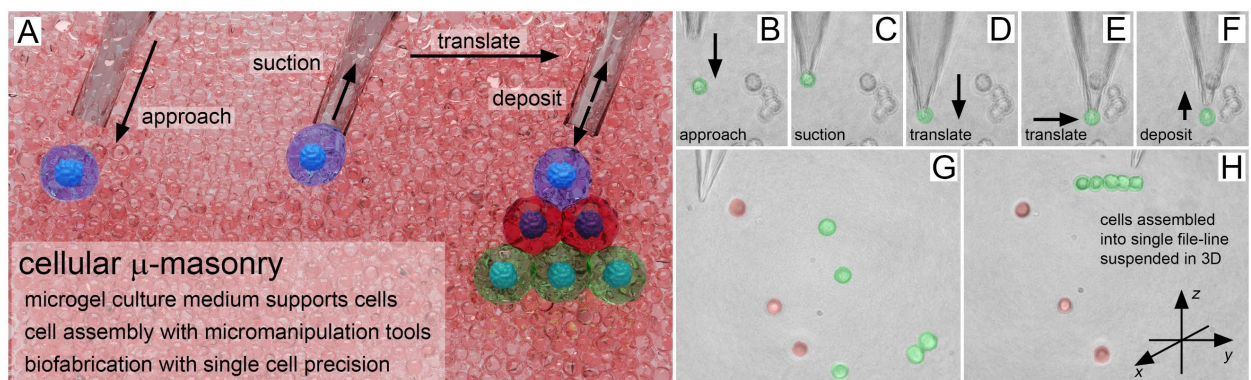

**Fig. 1.** (A) With the micromasonry technique, cellular structures are precisely assembled in 3D space within a microgel-based culture medium that provides stabilizing support. Dispersed cells are retrieved, translated, and deposited using a micromanipulator equipped with a glass microcapillary connected to a suction generator, all mounted on a confocal microscope. Shown here, a 3t3 fibroblast cell suspended in 3D is (B) approached and (C) retrieved by applying a light suction. (D,E) The cell is translated to a chosen location in 3D space and (F) deposited. (G) Cells are dispersed in 3D; green cells are selected for assembly, while red cells are left in place (3t3 fibroblasts, false coloring). (H) The green cells are picked up, translated, and placed next to each other forming a single-file line of cells, suspended in 3D space.

To provide support to cells while minimizing the physical shear stress they endure as we build structures, ensuring the cells are gently cradled in their 3D microenvironment, we optimize the microgel medium through rheological testing. We find that microgel media formulated at 5 - 6 wt% polymer has an elastic shear modulus of 10 - 20 Pa and a yield stress of 1.3 - 2.4 Pa (see Fig. S1 and Methods for microgel polymer species). This level of shear stress is comparable to experimental fluid stresses typically imposed on cell surfaces, so we do not expect this procedure to lead to cell damage *(19)*. As the microcapillary translates back and forth throughout the micromasonry process of building structures, the microgel "balls" are forced to rearrange and flow around the microcapillary surface. To determine whether these rearrangements lead to irreversible, long ranged, or unpredictable flow patterns in the microgel medium, we perform video imaging of a microcapillary translating through the microgel medium containing dispersed cells, moving at approximately 0.5 mm/s, which is the rate we translate the microcapillary during micromasonry procedures. We find that when the microcapillary is reciprocally translated near suspended cells, the net cell displacement is approximately one cell diameter or less; cells further from the capillary exhibit less hysteresis than cells directly in the path of the microcapillary (Fig S2).

Quantitative analysis of the microgel flow-field around the translating microcapillary helps to explain this reversibility in microgel displacement during the micromasonry process (Fig. S3). To understand this apparent hydrodynamic reversibility of the micromasonry method, we estimate the Reynolds' number, $Re$, given by $\rho v d / \eta$, where $\rho$ is the microgel mass density, $v$ and $d$ are the microcapillary translation speed and diameter, respectively, and $\eta$ is the medium viscosity *(20)*. The microcapillary diameter, $d$, is 1 mm along its shaft and approximately 5 μm near its tip, so we approximate the shear-rate range to be $v / d \approx 0.5 - 100$ s$^{-1}$. The corresponding microgel viscosity range from rheological measurements is 0.2 - 10 Pa s (Fig S1C). Thus, we estimate the maximum $Re$ occurring during micromasonry to be approximately $10^{-3}$, four orders of magnitude below the flow regime where hydrodynamic reversibility begins to break down *(21)*. Consequently, the predictable flow behavior of the packed microgel medium enables the assembly of precise structures in 3D space, like single-file lines of cells (Fig. 1G,H).

The heterogeneous composition of tissues *in vivo* often exhibit cell-type variations over length scales as small as the individual cell; to mimic this extreme variability *in vitro*, we build structures from cells labeled with different fluorescent dyes. Two populations of Madin-Darby Canine Kidney (MDCK) epithelial cells are cultured under standard 2D conditions, labeling one population with CellMask orange and the other with 5-chloromethylfluorescein diacetate (CMFDA). To create a source population of cells to build with, the cells are harvested and suspended in a glass-bottom petri dish filled with 2 mL of the microgel culture medium. The dish is placed on an inverted confocal microscope equipped with an incubating plate, keeping the cells at 37 C (Fig 2A). To test our ability to generate a diversity of different spatial patterns that may occur in different tissues, we assemble several basic structures: a line of alternating colors, a triangle with rows of alternating color, a six-fold packing of red cells around a central green cell, and the unit cell of a honeycomb lattice (Fig 2B). We envision using micromasonry to study the emergence of collective behavior as a function of tissue size. To demonstrate this capability, we built different sized structures of the same repeating checkerboard pattern (Fig. 2C). While some highly ordered tissues exhibit regular patterns like those shown here, even these highly ordered structures often lay on curved manifolds in space *(22)*. For example, the organ of Corti exhibits highly ordered checkerboard patterning and hexagonal packing over the curved surface of the cochlear duct *(22)*. Thus, to explore the possibility of using micromasonry to build larger objects with larger-scale structural complexity than simple geometric shapes, we constructed the initials of our institution, 'UF' out of almost 100 cells, suspended in 3D space (Fig 2D). Taken together, these tests of patterning at the single-cell scale and complex structuring at the large-scale demonstrate the potential for using micromasonry to mimic the complexity of *in vivo* tissues.

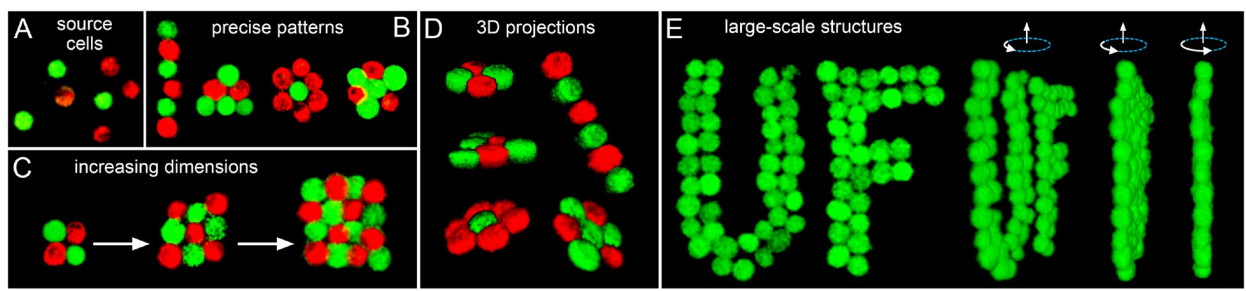

**Fig. 2** Two different MDCK cell populations are assembled into precise structures. (A) Red and green cells are dispersed randomly in the microgel medium. (B) Dispersed cells are retrieved and assembled into patterns with single cell precision, like the single-file line with alternating cell colors, shown here. While hexagonal packing is expected for spheres, with micro-masonry such structures can be made with different patterning. (C) Square packings are possible and the emergence of collective behavior can be studied by increasing the dimensions of a given pattern. (D) The small-scale patterns are shown from different angles to demonstrate control over cell placement in the third dimension. The cells are seen to be co-planar, suspended in 3D space. (E) Single cells are arranged into large irregular shapes of specific design, such as the initials of the authors' institution. Viewed from different angles, we see that cell placement along the third dimension is extremely precise.

To determine if the cells in these fabricated structures are functionally interacting with one another, we use a calcein dye assay to test whether gap junctions form. Gap junctions are plaques of intercellular nano-channels that form between neighboring cells that allow the diffusion of small molecules from cell to cell. This transport can be visualized using calcein acetoxymethyl (AM) ester (calcein AM), a cell-permeant live cell dye *(23)*. In live cells, calcein AM is converted to freely diffusing green fluorescent calcein through acetoxymethyl ester hydrolysis, intracellularly. When gap junctions are present, calcein dye can be observed passing from cell to cell with fluorescence microscopy *(23)*. We culture MDCK cells in 2D, dye separate populations with CellMask orange and calcein AM, harvest the cells, and randomly disperse both populations as described above. Using the micromasonry technique, we build single-file lines of four red cells and then add a single green cell to the end of the line (Figure 3A). A confocal Z-stack is taken every 30 minutes for 24 hours. Throughout this period of time we see the green dye travel down the line of cells, indicating that gap junctions indeed form in these manufactured cellular structures (Figure 3B). We find that the calcein dye takes about 5 hours to travel from cell to cell, which agrees with results from standard calcein assays *(23)*. To ensure this observation requires gap junction permeable fluorophores, we perform control experiments in which CMFDA is used in place of calcein; CMFDA cannot pass through gap junctions. In these control experiments, the green dye does not spread from cell to cell (Figure 3A). These results demonstrate that structures suspended in the 3D microgel microenvironment, assembled with the micromasonry technique, maintain their capacity to form the functional gap junctions typically observed in more traditional culture contexts.

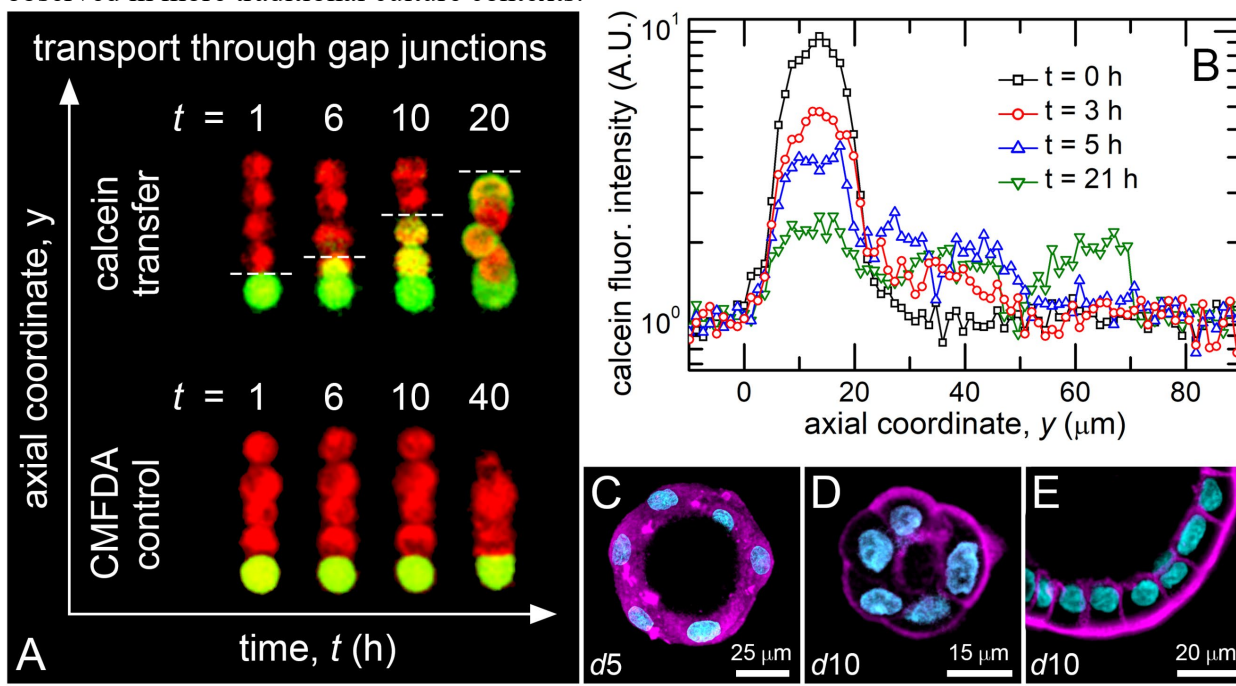

**Fig. 3** Functional assays. (A) As a functional test, we perform a calcein transport assay. Four red cells are assembled in a line and one cell dyed with calcein is placed on the end (top). All the cells become fluorescent green over 20 hours, indicating that gap junctions form, allowing the calcein dye to travel from cell to cell. In control experiments, a cell dyed with CMFDA is placed on the end of the red-cell line; CMFDA is gap-junction impermeable (bottom). We see no CMFDA transport in these experiments. (B) Space-time analysis of calcein fluorescence intensity shows the calcein transferring from the source cell to the neighboring cells. (C, D, E) As a second functional assay we study acini formation in the microgel medium. (C) After five days we see hollow shells forming with disordered cytoskeletal structure. (D,E) After 10 days we see structures resembling mature acini (cyan: Hoescht; magenta: Alexa-phalloidin).

As a second functional assay, we test whether glandular acini can form within microgel media. Acini represent one of the best established and widely used tissue models; acini formed from mammary epithelial cells are used in breast cancer research, for example *(7-10)*. Paralleling standard protocols, we disperse MDCK cells into a modified microgel media that is swollen in diluted Matrigel (see Methods). After about 10 days of incubation, we see that single cells have proliferated and self-assembled into the monolayer shell structures characteristic of traditionally cultured acini. To compare the architecture of these epithelial shells to traditional acini, we fix and stain them with Hoeschst 33342 and Alexa 594 phalloidin to visualize the nucleus and actin cytoskeleton. The stained tissues are then imaged with confocal microscopy where we see the characteristic monolayer shell structure as well as the polarized cytoskeletal structure typically found in acini (Fig. 3C-E). Slices through the 3D confocal stacks exhibit actin assembly near the outer-facing surface of the shell where a basement membrane is known to form *(7-10)*. These results represent a new way to culture glandular acini and point toward a future path of rich exploration; the cells' mechanical microenvironment can be tuned by preparing the microgel medium at different concentrations, and the granular nature of the microgel medium allows for the micromasonry technique to be combined with spontaneous acini formation. For example, different cell types can be delivered to the maturing acini at chosen locations and times, or concentrated doses of growth factors or other stimulatory molecules can be locally perfused with the micromasonry instrument. Similarly, such a hybrid approach could be used to expand the experimental toolbox for broader investigation of diverse tissue models of healthy development or disease processes *(24, 25)*.

To take the first steps toward using micromasonry for studying more complex cellular structures and potentially manipulating their function, we pack pluripotent embryonic cells around a functionalized microsphere. Following established protocols, blastula stage embryos (24 hours post fertilization) from the starlet sea anemone, *Nematostella vectensis*, injected with mRNA for green fluorescent protein as zygotes, are dissociated into single cells in calcium and magnesium free sea water. The dissociated cells are manually collected with a micropipette and dispersed in the microgel medium. Since these embryos are cultured in sea water, we developed a zwitterionic microgel formulation that does not de-swell at high salt concentrations (see Methods). We first built several simple planar structures from the dissociated cells (Fig. 4A). We then dispersed fibronectin-coated polystyrene microspheres in the microgel medium and built a hybrid biotic/abiotic structure in which the dissociated embryonic cells were deposited on the bead's surface (Fig. 4B). Time-lapse microscopy revealed the cells remained viable and motile, actively spreading on the bead over the course of 13 hours. While further investigation is needed to study how these cells respond to a process of disassembly and re-assembly on a foreign surface, this first demonstration of hybrid biotic/abiotic assembly is key to developing advanced biomaterials that precisely combine living cells with engineered microstructures *(26)*.

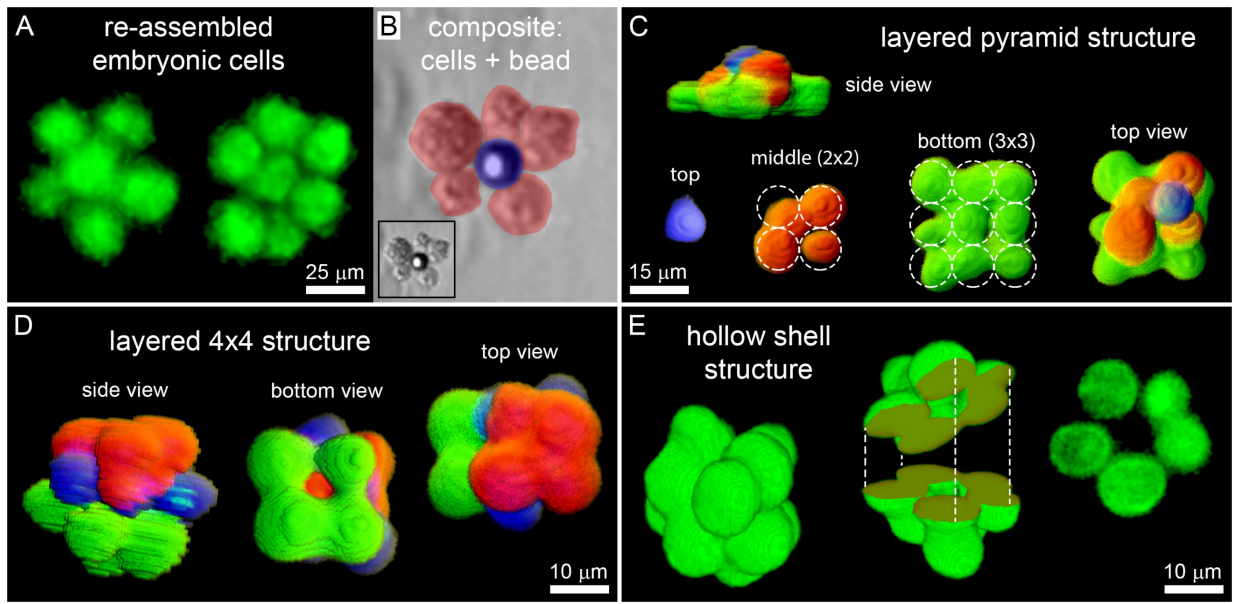

**Fig. 4** (A) Fluorescently labeled cells (GFP) from the sea anemone *Nematostella vectensis* embryos are disassociated, dispersed in the microgel medium, and used to build precise planar structures. (B) The disassociated embryonic cells are placed around a fibronectin-coated bead to observe their interactions with an anchoring surface in 3D (false colored; inset: uncolored bright-field image). (C) MDCK cells are dyed blue, red, and green to create three different populations. Three-color, 3D structures are assembled including a layered pyramid and (D) a layered stack of 2x2 structures. (E) A hollow spherical structure is built from MDCK cells dyed with CMFDA mimicking the structure of an acinus (left and center: volume-view renderings; right: X-Y slice).

To test for the potential of using micromasonry to build layered 3D structures having heterogeneities over length-scales of single cells, approximating the level of detail found in dense living tissues, we build a series of stratified objects from three separate cell populations. Extending the methods described above, we label MDCK cells with blue, green, and red dyes and selectively retrieve chosen cells from a randomly dispersed population to build layered patterns. We create a square pyramid structure by assembling a planar 3x3 square packing of green cells, followed by a 2x2 layer of red cells, finishing with a single blue cell at the apex (Fig. 4C). Similarly, we build a stack of 2x2 layers, with green cells forming the base, blue cells forming the middle layer, and red cells on top (Fig. 4D). Although there are imperfections in both structures, no imperfection is more than a single cell diameter, indicating that extremely precise, multicellular structures can be built using this method. As a final test of the potential for using micromasonry to build tissue models, we assemble a spherical shell of MDCK cells to approximate a glandular acinus. 3D renderings of this assembly reveal its shell structure; slices through the 3D structure show the open pore-space inside the shell of cells (Fig. 4E). While these structures were imaged immediately after building, our functional assays of MDCK structures showing gap junction formation and the development of characteristic acini structure over time indicate that these more complex 3D structures may evolve into functional tissue models (Fig. 3). We envision that the combination of micromasonry and optimal culture conditions will facilitate building "acini on demand" where the rapid integration of structural and microenvironmental cues could accelerate the development of mature acini, with the possibility of extending this principle to other tissue models.

**Conclusion**

All the cellular structures shown here were assembled by operating the micromasonry system by hand, in which a researcher dispersed cells into the microgel medium with a pipette, identified individual cells by eye on a microscope, and meticulously assembled them into the targeted designs by turning dials on micromanipulator control hardware. While single-cell precision was achieved with this manual approach, the current procedure limits the physical scale of structures that can be assembled. However, we believe that all the steps in cellular micromasonry can be automated by combining a diversity of current engineering tools like 3D image segmentation, 3D cell tracking, and the control algorithms of pick-and-place robotics *(27)*. Since building within the microgel medium creates uncertainties like elastic and plastic material deformations, we believe new computational tools based on machine learning and artificial intelligence for controlling machine operations in uncertain environments can be employed as a path forward to rapidly build larger and more complex structures with an even higher level of precision than that demonstrated here *(27)*.

As improved micromasonry tools are developed for creating larger-scale structures, smaller-scale manual micromasonry can now be used to explore the structure-function relationship in model tissues. Cell to cell signaling occurs extensively throughout embryonic development, with responding cells changing their cell fate and behavior in response to embryonic "organizing centers". In some embryos these organizing cells can be reduced to a single identifiable cell that establishes the fate of all surrounding cells *(28)*. By dissociating developing embryos and using the micromasonry technique to deliver the organizing cell to different complements and spatial distributions of surrounding "responding" cells, the sensitivity of cell signaling to cell positioning in these processes can be quantitatively investigated for the first time. Furthermore, the ability to easily manipulate gene expression (e.g. ligands, receptors, extracellular matrix associated molecules, transcription factors) in model developmental systems such as *Nematostella* provide powerful opportunities to understand the biology and biophysics of cell-cell interactions *(29)*. Similarly, a hybrid approach to organoid engineering could be developed. A dominating paradigm in organoid research is to program successive stages of differentiation into pluripotent cells that will spontaneously and collectively mature into functioning differentiated cellular structures that approximate mature organ behavior *(30)*. By building pre-structured assemblies from pluripotent cells and delivering additional programmed cells at precisely chosen locations at critical time-points, the organoid maturation process could be rapidly and controllably guided down many steps of differentiation and development. Finally, with the ability to create cellular structures having crystalline symmetry and spacing, as shown in Figure 2, the tools of synthetic biology could be used to create collective phases of cellular behavior that cannot be achieved within randomly structured cell assemblies. For example, cells can be programmed to secrete molecules at a periodic rate, introducing the possibility of coupling temporal oscillations of signaling molecule concentration with the spatial frequency of cell location. These "cell crystals" could be assembled with the microsmasonry method to synchronize their behaviors with a level of coherence and function that would be destroyed by random cell patterning. While current trends in 3D bioprinting research largely focus on fabricating large-scale functional engineered tissues, we hope the cellular micromasonry technique introduced here inspires researchers to also work in the opposite direction, using the methods, materials, and tools of biofabrication to conduct fundamental investigations of collective cell behavior at a level of structural detail approaching that found *in vivo*.

**Supplementary Materials:**

Materials and Methods

Figures S1-S4

Movies S1-S2

**Materials and Methods**

**Jammed Microgel Synthesis**

To synthesize polyacrylamide microgels with 17 mol% methacrylic acid, a solution of 8% (w/w) acrylamide, 2% (w/w) methacrylic acid, 1% (w/w) poly(ethylene glycol) diacrylate (MW = 700 g mol−1), and 0.1% (w/w) azobisisobutyronitrile in ethanol (490 mL) is prepared. The solution is sparged with nitrogen for 30 min, then placed into a preheated oil bath set at 60 °C. After approximately 30 min, the solution becomes hazy and a white precipitate begins to form. The reaction mixture is heated for an additional 4 hours. At this time, the precipitate is collected by vacuum filtration and rinsed with ethanol on the filter. The microparticles are triturated with 500 mL of ethanol overnight. The solids are again collected by vacuum filtration and dried on the filter for approximately 10 min. The particles are dried completely in a vacuum oven set at 50 °C to yield a loose white powder. The purified microgel powder is dispersed in cell growth media at various concentrations and mixed at 3500 rpm in a centrifugal speed mixer in 5-min intervals until no aggregates are apparent (17, 18). The microgel is then neutralized to a pH of 7.4 with NaOH and 25mM HEPES buffer (Part no. BP299-100) and is left to swell overnight, yielding microgel growth media at concentrations of 5-6% (w/w).

In the case of embryonic experiments, zwitterionic microgels are synthesized in the same method from a solution of acrylamide (AAm), poly(ethylene glycol) diacrylate (PEGda) (MW = 700 g mol$^{-1}$), azobisisobutyronitrile (AIBN), and carboxybetaine methacrylate (CBMA) as an ionisable comonomer in ethanol (490 mL). The purified microgel powder is dispersed in sea water at 8% polymer concentration and mixed at 3500 rpm in a centrifugal speed mixer in five-minute intervals until no aggregates are apparent. The microgel is then left to swell overnight, yielding microgel the 3D growth media.

**Cell Culture**
MDCK cells (Madin Darby Canine Kidney epithelial cells, NBL-2 ATCC CCL-34) are cultured in Dulbecco's Modified Eagle Medium (DMEM) with 4.5 g/L glucose, L-glutamine, and sodium pyruvate supplemented with 10% FBS and 1% penicillin streptomycin in a 12 well plate. When the cells have reached 70% confluence, a single well is dyed with live cell dye. For control studies, one well is dyed with CellMask orange plasma membrane stain (Thermo-Fisher, part no. C10045) and a separate well with CellTracker green (CMFDA) (Thermo-Fisher, part no. C2925). For gap junction investigations, one well is dyed with CellMask orange and a separate well with CellTraceCalcein Green AM (Thermo-Fisher, part no. C34852), which becomes fluorescent green in live cells. We passage the wells separately by washing with PBS then incubating in 5% Trypsin—EDTA solution for 5 minutes. The cells are harvested from the plate and placed into separate 15mL centrifuge tubes, where they are centrifuged at 650g for 4 minutes. The supernatant is removed and each pellet is suspended in 1 mL of fresh cell growth media. The cell pellets are dispersed with gentle pipette mixing and 100 µL of each solution is placed in a fresh 15 mL tube and mixed together with light pipetting. About 50 µL of the solution is dispersed in the jammed microgel using a pipet. The microgel 3D culture medium is prepared for each cell type using the corresponding liquid media. 35-mm glass bottom petri dishes are used to facilitate fluorescence imaging. 3 mL of microgel media is loaded into each well. Prior to adding cells, dishes containing microgel media are incubated at 37 °C and 5% CO2.

**Embryo Experiments**
*Nemaotstella vectensis* embryos were obtained from a breeding colony in the Martindale lab at the Whitney Lab for Marine Bioscience (Univ. Florida). Zygotes are deljellied and injected with mRNAs to GFP or mCherry to generate fluorescently labeled blastomeres. Embryos are grown to blastula stages (24-48 hours post fertilization) and dissociated into individual cells using calcium/magnesium free seawater plus 5 mM egtazic acid (EGTA). The embryos are placed in a petri dish and approximately 2 mL of the dissociation solution is added. The embryos are mechanically dissociated using a pipette for 1 minute or until there are no large pieces of tissue visible. The dissociated embryo solution is transferred to a 40 µm cell strainer positioned over a clean petri dish and washed with the dilution solution.

**Building Single Cell Structures**

Once the cell loaded petri dish is prepared, it is moved to a Nikon C2+ laser scanning confocal microscope. An incubating plate is used to keep the dish warm during the building process. A thin layer of oil is placed over the gel to prevent evaporation. Microcapillary translation is achieved using a Siskiyou MX7600 micromanipulation system and the capture and release of cells is achieved using an Eppendorf CellTram. The microcapillaries are fabricated using a Sutter P-97 micropipette puller and smoothed using a Narshige microforge. To initiate the micromasonry process, the microcapillary is slowly lowered into the gel and centered in the chose field of view using a 4X objective. A 'layout' image is taken at 10X magnification to visualize cell locations and colors for specific builds. Structures are built using a 10X objective with an additional 1.5X zoom lens.

**Supplementary Text**

Rheology of Jammed Microgels

The methacrylic acid microgel particles are synthesized in house, as described in Materials and Methods, and swollen in DMEM supplemented with 5% FBS and 1% penicillin streptomycin, providing the necessary nutrients for cells to function. To determine the material and flow properties of the jammed microgels, we conduct traditional rheological tests including frequency sweeps (1% strain amplitude) and unidirectional shear-rate sweeps (Fig. S1). We find that microgel media prepared within a polymer concentration range of 5 - 6% is well suited for use with the micromasonry technique. Within this range, the microgel media has an elastic shear modulus of 10 to 20 Pa and a yield stress of 1.3 – 2.4 Pa. The stresses measured in shear-rate sweeps are often divided by the shear-rate to determine the viscosity. While the microgel media is dominantly solid-like at low shear-rates, at high shear rates the microgel media becomes a shear-thinning fluid, exhibiting a decreasing viscosity with increasing shear-rate. We use this viscosity curve to estimate the range of Reynold's number the media exhibits during micromasonry procedures.

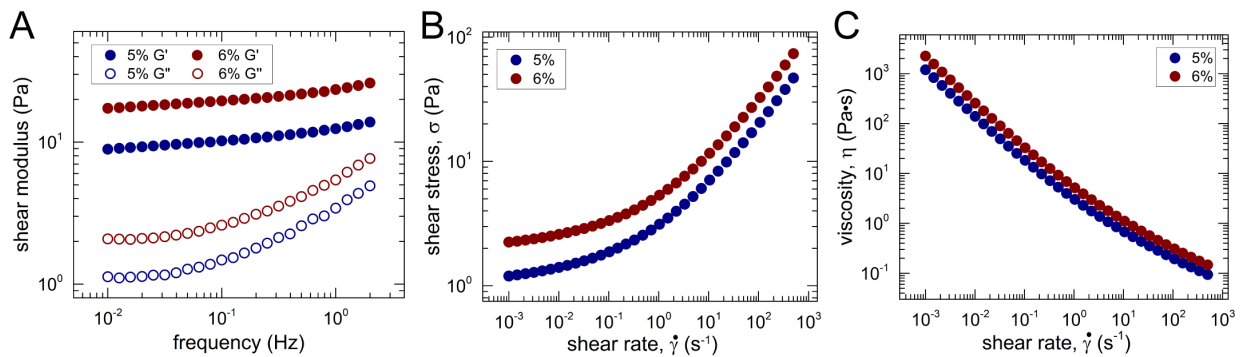

**Fig. S1.** A) A frequency sweep is performed from $10^{-2}$ to 2 Hz. $G'$ is weakly dependent on frequency and always greater than $G''$ for both 5% and 6% MAA microgels. These rheological properties indicate that the microgel media is dominantly solid-like at low levels of strain. B) To determine the yield stress of the microgel media, we perform shear-rate sweeps while measuring shear stress. The plateau in shear stress at low shear-rate corresponds to the yield stress of the microgel media, which we find to be between approximately 1.3 and 2.4 Pa. C) While the viscosity of the microgel media is several orders of magnitude greater than water, cells experience extremely low shear stresses during micromasonry procedures because of the low shear-rates involved.

Microgel Flow and Reversible Cell Displacement Near Translating Capillary Tips
During micromasonry procedures, it is possible that the microgel medium flows excessively and that dispersed cells displace irreversibly because of material's rheological properties. To investigate the degree of microgel flow and irreversible cell displacement during micromasonry procedures, we performed a series of tests in which a microcapillary is translated back and forth near suspended cells. Bright-field and fluorescence videos are collected throughout the tests to monitor the motion of the microcapillary, the cells, and the supporting microgel medium (Figure S2). After one cycle, we see that the cells displace by less than one cell diameter. Since the microgel medium is inherently an optically grainy material, we can use particle image velocimetry (PIV) to measure the entire displacement field after one cycle (Figure S3). We find the root-mean-square displacement, averaged over the entire field of view, to be 4 μm, which is less than a single cell diameter.

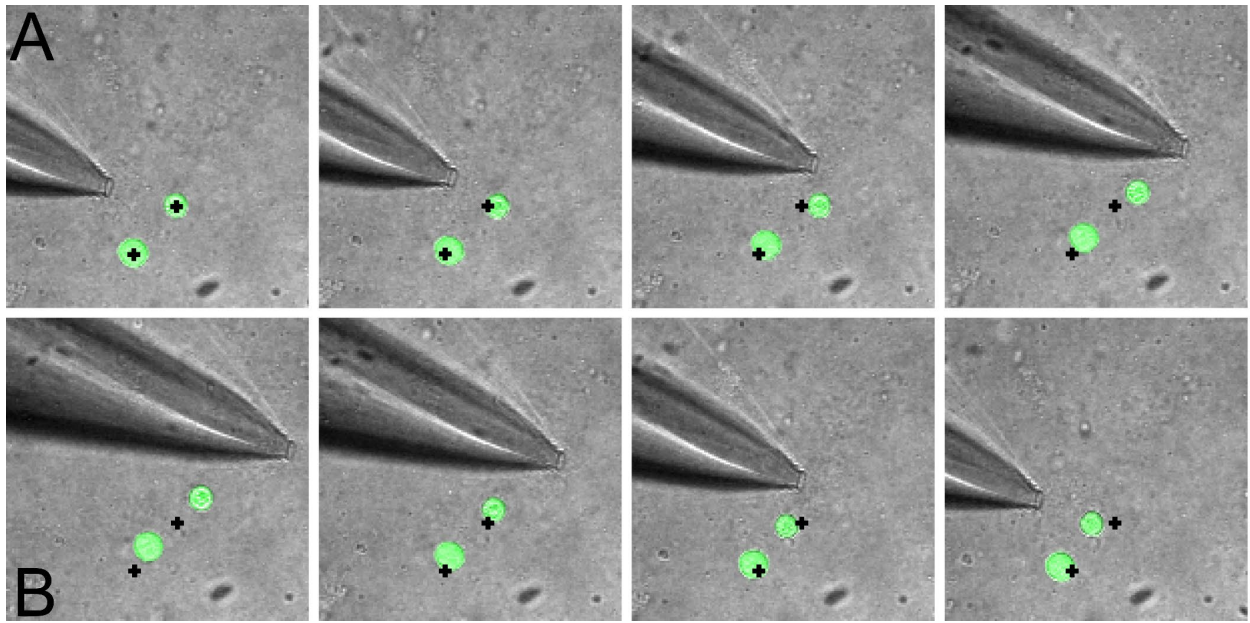

**Fig. S2.** MDCK cells are dyed with cell tracker green and dispersed in 5% MAA microgel media. (A) A glass microcapillary is translated rightward, passing two cells. The cells' initial positions are marked with a "+". The cell closest to the capillary displaces by more than 1 cell diameter; the cell further from the capillary displaces by less than one diameter. (B) The needle tip is translated leftward, back past the cells. Both cells are moved toward their initial locations, having a final displacement of less than one cell diameter.

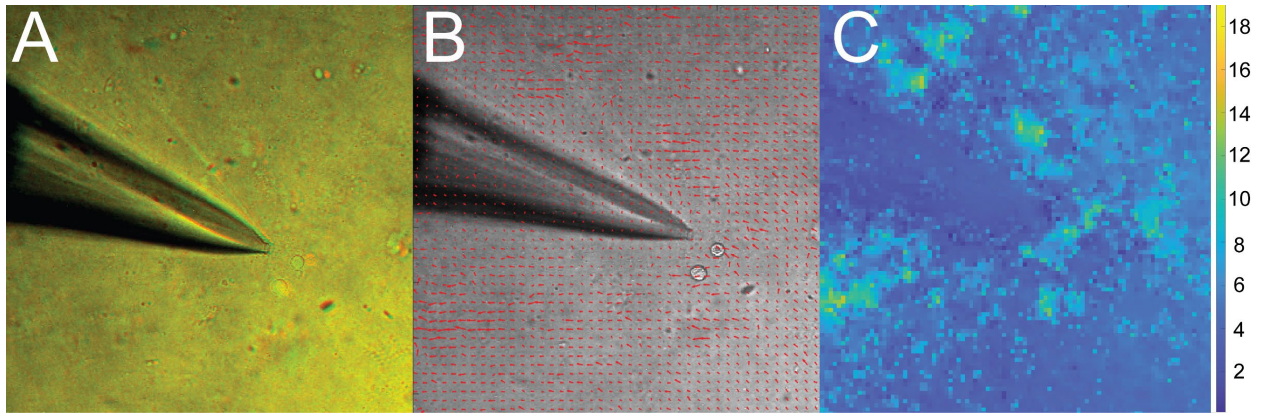

**Fig. S3.** The microcapillary is translated right and left, past two MDCK cells, through 5% MAA microgel media. (A) False coloring the initial time-point in green and the final time-point in red, an overlay image shows the net displacements through a visible red and green patchiness. (B) We perform PIV to quantify the displacement field throughout the field of view. (C) Creating a displacement map from the absolute values of the displacement vectors, we see that displacements are patchy and largest near the microcapillary. The RMS displacement is 4 μm.